\documentclass{article} 
\usepackage{geometry}
\usepackage[numbers]{natbib}
 \usepackage{subfigure}
 \usepackage{booktabs}
 \usepackage[table,xcdraw]{xcolor}
 \usepackage{mdframed,lipsum}
\usepackage[english]{babel} 
\usepackage{xcolor}
\usepackage{listings}
\usepackage{amssymb}
\usepackage{amsmath}
\usepackage{multirow}
\usepackage{txfonts}
\usepackage{multicol}
\usepackage{booktabs}
\usepackage{mathdots}
\usepackage{graphicx} 
\usepackage{fancyhdr}
\usepackage{hyperref}
\usepackage{float}
\usepackage{microtype}
 \usepackage{textcomp}
\usepackage{caption}
\newtheorem{definition}{Definition}
\definecolor{ao(english)}{rgb}{0.0, 0.5, 0.0}
\definecolor{bostonuniversityred}{rgb}{0.8, 0.0, 0.0}
\definecolor{antiquebrass}{rgb}{0.8, 0.58, 0.46}
\definecolor{antiquefuchsia}{rgb}{0.57, 0.36, 0.51}
\definecolor{asparagus}{rgb}{0.53, 0.66, 0.42}
\definecolor{cornsilk}{rgb}{1.0, 0.97, 0.86}
\usepackage{tikz}
\usetikzlibrary{%
	arrows,
	calc
}
\makeatletter
\def\@xfootnote[#1]{%
  \protected@xdef\@thefnmark{#1}%
  \@footnotemark\@footnotetext}
\makeatother

\usepackage{sectsty}
\subsectionfont{\normalfont\itshape}
\subsubsectionfont{\normalfont\itshape}

\begin{document}


\begin{tabular}{p{1.1in}p{4.5in}p{1.2in}}  
\hspace{-1cm}
\noindent
\begin{tabular}{c}  \includegraphics[width=2.9cm]{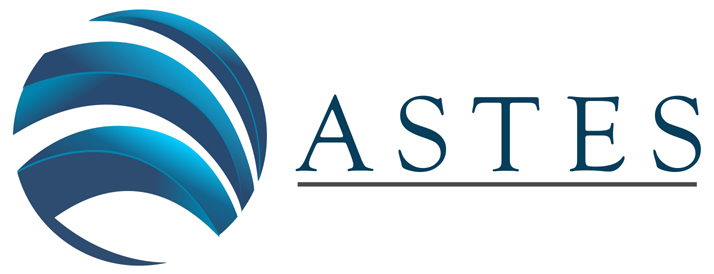}\end{tabular} 	& \vspace{-0.5cm} \centering \textit{Advances in Science, Technology and Engineering Systems Journal \newline Vol. 5, No. 5, 1189-1195 (2020)} \\   \href{http://www.astesj.com}{www.astesj.com}  
	& \vspace{-0.6cm}  \rule{1.2in}{1.5pt} \vspace{-0.2cm} \newline \centering \textbf{ ASTES Journal \newline ISSN: 2415-6698} \newline \rule{1.2in}{1.7pt} 
\end{tabular}

\vspace{1.8cm}

\noindent \textbf{\LARGE{\setlength\itemsep{0pt}Synthesis of SQL Queries from South African Local Language Narrations}}

\vspace{0.2cm}

George Obaido\footnote[*]{Corresponding Author: George Obaido, University of the Witwatersrand,\\ School of Computer Science and Applied Mathematics, Johannesburg, 2001, South Africa. \\Email: rabeshi.george@gmail.com \& Contact: +27832227427 }${}^{,12}$, Abejide Ade-Ibijola${}^{2}$, Hima Vadapalli${}^{1}$ 

\vspace{0.2cm}
 \textit{${}^{1}$School of Computer Science and Applied Mathematics, University of the Witwatersrand, Johannesburg, 2001, South Africa.}

\vspace{0.2cm}
\textit{${}^{2}$School of Consumer Intelligence and Information Systems, University of Johannesburg, Johannesburg, 2001, South Africa.}

\vspace{0.3cm}

\begin{tabular}{p{1.7in} p{0.1in} p{5.0in} }
A R T I C L E \hspace{0.1cm} I N F O &  & A B S T R A C T \\ 
 \cline{1-1}  \cline{3-3} \setlength\itemsep{0pt} \vspace{-0.1cm}
\textit{Article history:
	\newline Received: 26 August 2020
	\newline Accepted: 18 October 2020
	\newline Online: 24 October 2020
	\newline \rule{1.78in}{0.5pt} 
	Keywords: 
	\newline Language translation
	\newline Synthesis of things
	\newline JFA applications
	\newline SQL queries
}
 \newline \newline  & & \vspace{-0.1cm} 
 \textit{English remains the language of choice for database courses and widely used for instruction in nearly all South African universities, and also in many other countries. Novice programmers of native origins are mostly taught Structured Query Language (SQL) through English as the medium of instruction.  Consequently, this creates a myriad of problems in understanding the syntax of SQL as most native learners are not too proficient in English. This could affect a learner's ability in comprehending SQL syntaxes. To resolve this problem, this work proposes a tool called local language narrations to SQL (\texttt{Local-Nar-SQL}) that uses a type of Finite Machine, such as a Jumping Finite Automaton to translate local language narratives into SQL queries. Further, the generated query extracts information from a sample database and presents an output to the learner. This paper is an extension of work originally presented in a previous study in this field. A survey involving 145 participants concluded that the majority found \texttt{Local-Nar-SQL} to be helpful in understanding SQL queries from local languages. If the proposed tool is used as a learning aid, native learners will find it easier to work with SQL, which will eliminate many of the barriers faced with English proficiencies in programming pedagogies.}\\
 \cline{1-1}  \cline{3-3}
\end{tabular}

\vspace{0.5cm}

\begin{multicols}{2}

\section{INTRODUCTION}
According to Linguists, English has much of its vocabulary borrowed from French and Latin, but it remains a Germanic language in terms of structure and sounds~\cite{fulk2008english,bech2016english,borodenko2019formation}.
English is the predominant language of instruction used at most universities and remains the \textit{de facto} language of communication for many industry sectors~\cite{ndhlovu2018english,passera2017diagrams}. It is interesting to note that many of the world's population do not speak English as their first language, but it is vastly used for communication and dissemination of knowledge~\cite{inbar2010english,guo2018non}. Most programming and scripting languages, such as C, Java, Python and Structured Query Language (SQL) have keywords, declarations and documentations based in English~\cite{guo2018non,reestman2019native,soosai2019effect}. Even online resources, such as forums, blogs and developers communities that aid the learning of programming are primarily in English~\cite{nguyen2018statistical,barua2014developers}. For example, Stack Overflow, which is one of the largest online communities for developers, enforces that all questions and answers are posted in English. Hence, for a novice programmer to be proficient in programming, English is a necessity~\cite{reestman2019native}.

SQL is the most popular query declarative language used for performing operations on relational databases~\cite{obaido2019synthesis,ade2017s}. As a standardised language, many of its syntaxes and statements are English-like, and SQL has found numerous applications in both industry and academia~\cite{obaido2018generating,obaido2019generating}. Listing \ref{lst:label1} shows examples of the SQL SELECT query statements that depict English-like formation. These set of queries would retrieve one or more records from a targeted database's table.

\begin{lstlisting}[language=SQL,escapechar=@, morekeywords={TRUNCATE},frame=bt, numbers=none, label={lst:label1}, caption={SQL SELECT statement query }]
SELECT * FROM myrecordinfo;

SELECT DISTINCT firstname, lastname
FROM myrecordinfo; 

SELECT * FROM myrecordinfo
WHERE firstname='karabo' AND lastname='hlophe';

SELECT * FROM Customers 
WHERE country='South Africa' 
OR city = 'Pietermaritzburg';
\end{lstlisting}

South Africa is a multilingual country with eleven official languages~\cite{brookes2020youth,coetzeebeing,banda2000dilemma}. Most South African learners prefer to communicate in their local dialects, rather than English~\cite{coetzeebeing,banda2000dilemma}. Educational instruction for learning programming requires a good proficiency in English, which creates numerous challenges for a local student whose native language is not English~\cite{guo2018non,pal2015classroom,veerasamy2014teaching}. Just like any other programming language, anecdotal evidence has shown that SQL is hard for students~\cite{obaido2019synthesis,ade2017s,obaido2018generating,migler2020mapping}. Even the straightforward, English-like syntax of SQL is misleading~\cite{prabhu2019sql,ardeleanu2016sql}. Other studies have identified that the burden of remembering SQL keywords and database schemas are often challenging~\cite{migler2020mapping,heller2019use}. These challenges have made it crucial for researchers and developers to create \textit{culturally-agnostic} aids that would assist learners to improve their understanding of SQL. Such systems could be developed to enable learners express their requests in \textit{free-forms} using local language narrations. To our knowledge, this is the first study that attempts to translate South African local language narrations into SQL queries.

\begin{figure}[H]
	\centering
	\includegraphics[width=\linewidth]{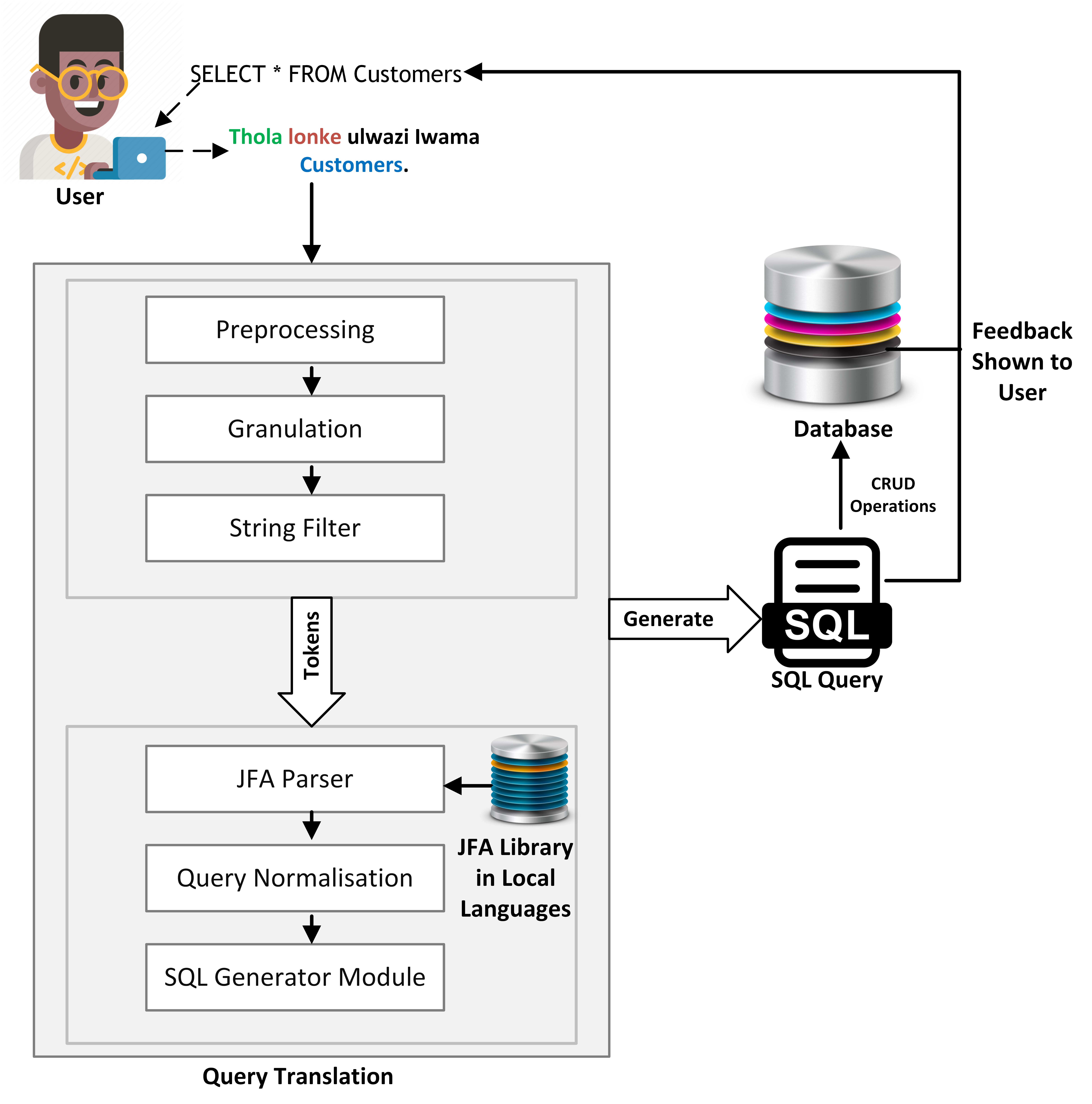}
	\caption{The translation process (adapted from \cite{obaido2019synthesis}) }
	\label{framework}
\end{figure}

This paper is an extension of work originally presented in \cite{obaido2019synthesis}, where natural language specifications were translated into SQL queries. We extended the use of a Jumping Finite Automaton, a type of Finite Machine for the translation of local language specifications into SQL queries. This idea was further implemented into a desktop-based application called \texttt{Local-Nar-SQL}. The process of translating the local language specifications into SQL queries is presented in Figure \ref{framework}.  Here, a user provides narrations in a local language, that is then processed by \texttt{Local-Nar-SQL}. The system preprocesses and passes the provided narrations to the JFA parser for matches. Further, the tokens are matched and a query is generated, which is made available to the user. In this paper, the following contributions were made. This study has:

\begin{enumerate}
	\item developed a method for the recognition of local language specifications using a JFA,
	\item extended this approach into a tool, called \texttt{Local-Nar-SQL}, for the translation of local language specifications into SQL queries, and
	\item evaluated this approach using human subjects and presented initial results.
\end{enumerate}
The rest of this paper is organised as follows: Section \ref{background} highlights the background and related works, Section \ref{methodology} presents the methodology used for this study, Section \ref{results} shows the implementation and results details, the evaluation information is provided in Section \ref{evaluation} and Section \ref{conclusion} concludes the paper and highlights discussions for future studies.

\section{BACKGROUND}\label{background}
This section presents the problems as research questions, highlights the motivation behind this work and discusses related work.
\subsection{Research Questions}
This study attempts to solve the following problems:

\begin{enumerate}
	\item how can we successfully translate local language narrations into SQL queries?
	\item can we describe a JFA and implement this into a tool that addresses the above problem? 
	\item what are human perceptions of the developed tool? 
\end{enumerate}
These problems were addressed in Section \ref{methodology}, \ref{results} and \ref{evaluation}.

\subsection{Motivation}\label{motivation}

\begin{description}
	\sloppypar
	\item[English proficiency is a problem in South Africa] English proficiency remains a major concern for local students learning programming~\cite{reestman2019native}. Many programming language constructs, keywords, API methods and libraries are mostly written in English, which pose a great deal of difficulty for a local learner~\cite{guo2018non,pal2015classroom,veerasamy2014teaching}. 
	\item[Decolonisation of science] Issues regarding the decolonisation of the science curriculum is currently being debated in South Africa and the rest of the world~\cite{de2016decolonisation}. Many local learners agree that the legacies of apartheid (entrenched colonialism) are still evident in education sectors, built by western perspectives, which might impinge access to learning~\cite{mudaly2018decolonising,le2016decolonising}.  
	\item[Comprehension problems in SQL] Generally, the English-like syntax of SQL is misleading for learners~\cite{ade2017s,obaido2019generating}. Similarly, most query interfaces are built to report errors in English, which might be confusing for learners to decipher~\cite{llopis2013make,barzdins2016towards}.
	\item[Abstract nature of programming] From a novice learners' perspective, the process of learning a new programming language remains a daunting task~\cite{ade2015introducing,ade2016automatic}. Even so, the abstract nature of programming has resulted in high failure rates for Computer Science (CS) disciplines~\cite{konecki2014problems,ade2014abstracting}.
\end{description}
\subsection{Why Local Language Narrations}
In Section \ref{motivation}, it was seen that English language poses numerous difficulties for local language learners. An ideal approach would be to enable such learners to express their intention, free from natural language ambiguities -- \textit{in narrations}. Narrations have been shown to provide textual explanations for programming and scripting language paradigms~\cite{obaido2019synthesis,obaido2018generating,ade2014abstracting}. This practice allows learners to express their requests in \textit{free-forms}, ignoring programming language syntax. Such \textit{free-forms} allow learners to express their requests in their local language, free from English language ambiguities. For programming languages, especially in low-resource settings, such an approach has been tried in \cite{soosai2019effect,soosai2018does,varadarajan2020declarative}.

\subsection{Related Work and Terms}
This section presents similar work in the area of language translations and the applications of JFA.

\begin{description}
	\sloppypar
	\item[Language Translation] An evolutionary method for translating Portuguese language into SQL query~\cite{afonso2008evolutionary}, a pattern matching approach for translating Hindi language into SQL query~\cite{kumar2014translating}, Telugu language translation into SQL query~\cite{reddy2006dialogue}, translating Italian language to SQL query~\cite{siciliani2019italian}, translating Chinese language into SQL query~\cite{meng2001nchiql} and Arabic translation into SQL queries~\cite{al1999portable,otair2008arabic,bais2016independent}.
	
	\item[JFA Applications] Abstraction of frequently asked questions (FAQs) for comprehension purposes~\cite{okwunma2019automatic}, synthesising SQL queries from narratives~\cite{obaido2019synthesis} and the automatic comprehension of tweets~\cite{obare2019jumping}.
	
	\item[SQL Comprehension] Generating narratives of SQL queries~\cite{ade2017s}, generating narratives of nested SQL queries~\cite{obaido2019generating}, generating SQL queries from visual specifications~\cite{obaido2018generating} and synthesising SQL queries from verbal specifications \cite{obaidotalksql}.
\end{description}

The method used in this paper is a JFA, and definitions of this term are as follows.

\begin{definition}[A Jumping Finite Automaton (JFA)~\cite{meduna2012jumping}]
	A JFA is a five-tuple, $M$ = ($Q$, $\Sigma$, $R$, $s$, $F$) where:
	
	\begin{enumerate}
		\item $Q$ is a bounded set of states,
		\item $\Sigma$ is an input alphabet,
		\item $R$ is the bounded set of rules, where $py$ $\rightarrow$ $q$ ($p$,$q$ $\in$ $Q$, $y$$\in$$\Sigma$), 
		\item $s$$\in$$Q$ is the start state, and
		\item $F$$\subseteq$$Q$ is the final state.
	\end{enumerate}	
\end{definition}

\section{METHODOLOGY}\label{methodology}
This section presents the methodology for this study. To begin, local languages were abstracted into a JFA, then queries were normalised, where irrelevant details were removed. Next, a query was generated.
\subsection{Abstracting Local Languages to SQL}
For local language abstraction into a JFA, entities were identified with colours, such as query types ($\sum _\text{QT} = a_{x}$) in green, attributes ($\sum _\text{CT} = b_{y}$) in red and relations ($\sum _\text{ET} = c_{z}$) in blue. This information is presented in Table \ref{jfa}. The Chinook DB~\cite{schultheiss2007storing} was selected for our use-case with a corpus containing a total of 120 different local languages conversions. This database contains a typical digital music store, consisting of eleven tables and multiple records.  Some examples of the JFA abstractions are presented. \\

\bigskip
\bigskip

\begin{center}
		\captionof{table}{JFA symbols 1 - 30}
	\begin{tabular}{ |p{2.3cm}|p{3.0cm}|p{2.3cm}|}
		
		\hline
		
		\hline
		$a_x$ & $b_y$ & $c_z$\\
		\hline
		$a_0$  \color{ao(english)}{Khombisa}   & $b_0$  \color{bostonuniversityred}{EmployeeID}    & $c_0$  \color{blue}{Employee} \\
		$a_1$  \textcolor{ao(english)}{Ungitholele}   & $b_1$  \textcolor{bostonuniversityred}{LastName}    & $c_1$  \textcolor{blue}{Genre} \\
		$a_2$  \textcolor{ao(english)}{Ngitholela}   & $b_2$  \textcolor{bostonuniversityred}{FirstName}    & $c_2$  \textcolor{blue}{Customer} \\
		$a_3$  \textcolor{ao(english)}{Ukuthola}   & $b_3$  \textcolor{bostonuniversityred}{Title}    & $c_3$  \textcolor{blue}{Mediatype} \\
		$a_4$  \textcolor{ao(english)}{Usika}   & $b_4$  \textcolor{bostonuniversityred}{ReportTo}    & $c_4$  \textcolor{blue}{Track} \\
		$a_5$  \textcolor{ao(english)}{Fumana}   & $b_5$  \textcolor{bostonuniversityred}{Address}    & $c_5$  \textcolor{blue}{Invoice\_item} \\
		$a_6$  \textcolor{ao(english)}{Khetha}   & $b_6$  \textcolor{bostonuniversityred}{State}    & $c_6$  \textcolor{blue}{Invoice} \\
		$a_7$  \textcolor{ao(english)}{Enta}   & $b_7$  \textcolor{bostonuniversityred}{City}    & $c_7$  \textcolor{blue}{Phone} \\
		$a_8$  \textcolor{ao(english)}{Faka}   & $b_8$  \textcolor{bostonuniversityred}{PostalCode}    & $c_8$  \textcolor{blue}{Playlist\_track} \\
		$a_9$  \textcolor{ao(english)}{Susa}   & $b_9$  \textcolor{bostonuniversityred}{Fax}    & $c_9$  \textcolor{blue}{Album} \\
		$a_{10}$  \textcolor{ao(english)}{Funda}   & $b_{10}$  \textcolor{bostonuniversityred}{Country}    & $c_{10}$  \textcolor{blue}{Artist} \\
		$a_{11}$  \textcolor{ao(english)}{Wys}   & $b_{11}$  \textcolor{bostonuniversityred}{Email}    & $-$  \textcolor{blue}{} \\
		$a_{12}$  \textcolor{ao(english)}{Vind}   & $b_{12}$  \textcolor{bostonuniversityred}{CustomerID}    & $-$ \textcolor{blue}{} \\
		$a_{13}$  \textcolor{ao(english)}{Lys}   & $b_{13}$  \textcolor{bostonuniversityred}{SupportRepID}    & $-$  \textcolor{blue}{} \\
		$a_{14}$  \textcolor{ao(english)}{Kry}   & $b_{14}$  \textcolor{bostonuniversityred}{TrackID}    & $-$  \textcolor{blue}{} \\
		$a_{15}$  \textcolor{ao(english)}{Plaas}   & $b_{15}$  \textcolor{bostonuniversityred}{ArtistID}    & $-$ \textcolor{blue}{} \\
		$a_{16}$  \textcolor{ao(english)}{Opdateer}   & $b_{16}$  \textcolor{bostonuniversityred}{InvoiceID}    & $-$ \textcolor{blue}{} \\
		$a_{17}$  \textcolor{ao(english)}{Verwyder}   & $b_{17}$  \textcolor{bostonuniversityred}{MediaTypeID}    & $-$ \textcolor{blue}{} \\
		$a_{18}$  \textcolor{ao(english)}{Skep}   & $b_{18}$  \textcolor{bostonuniversityred}{InvoiceLineID}    & $-$ \textcolor{blue}{} \\
		$a_{19}$  \textcolor{ao(english)}{Lees}   & $b_{19}$  \textcolor{bostonuniversityred}{Name}    & $-$ \textcolor{blue}{} \\
		$a_{20}$  \textcolor{ao(english)}{Nghenisa}   & $b_{20}$  \textcolor{bostonuniversityred}{UnitPrice}    & $-$  \textcolor{blue}{} \\
		$a_{21}$  \textcolor{ao(english)}{Sula}   & $b_{21}$  \textcolor{bostonuniversityred}{Composer}    & $-$ \textcolor{blue}{} \\
		$a_{22}$  \textcolor{ao(english)}{Susa}   & $b_{22}$  \textcolor{bostonuniversityred}{Company}    & $-$ \textcolor{blue}{} \\
		$a_{23}$  \textcolor{ao(english)}{Engetela}   & $b_{23}$  \textcolor{bostonuniversityred}{Al $|$ Alle $|$ Alles}    & $-$  \textcolor{blue}{} \\
		$a_{24}$  \textcolor{ao(english)}{Tumbuluxa}   & $b_{24}$  \textcolor{bostonuniversityred}{Konke $|$ Yonke}    & $-$  \textcolor{blue}{} \\
		$a_{25}$  \textcolor{ao(english)}{Tlakusisa}   & $b_{25}$  \textcolor{bostonuniversityred}{Yothe $|$ Vhothe}    & $-$  \textcolor{blue}{} \\
		$a_{26}$  \textcolor{ao(english)}{Ulonga}   & $b_{26}$  $-$\textcolor{bostonuniversityred}{}    & $-$  \textcolor{blue}{} \\
		$a_{27}$  \textcolor{ao(english)}{Ubvisa}   & $b_{27}$  $-$\textcolor{bostonuniversityred}{}    & $-$  \textcolor{blue}{} \\
		$a_{28}$  \textcolor{ao(english)}{Udadzisa}   & $b_{28}$ $-$ \textcolor{bostonuniversityred}{}    & $-$  \textcolor{blue}{} \\
		$a_{29}$  \textcolor{ao(english)}{Utomola}   & $b_{29}$  $-$\textcolor{bostonuniversityred}{}    & $-$  \textcolor{blue}{} \\
		$a_{30}$  \textcolor{ao(english)}{Usika}   & $b_{30}$  $-$\textcolor{bostonuniversityred}{}    & $-$  \textcolor{blue}{} \\
		\hline
	\end{tabular}
	\label{jfa}
\end{center}

\begin{mdframed}[backgroundcolor=cornsilk!20]	
	\begin{itemize}
		\item Ek will \textcolor{bostonuniversityred}{al} die \textcolor{blue}{customer} 
		besonderhede \textcolor{ao(english)}{vind}
	\end{itemize}
\end{mdframed}  
Example 1 above shows a query request sentence in Afrikaans. The English translation for this sentence is: ``I want to \textcolor{ao(english)}{find} \textcolor{bostonuniversityred}{all} the \textcolor{blue}{customer}'s details". The equivalent JFA shows:

\begin{center}
	\textbf{M} = $( \lbrace \textcolor{blue}{I, J, K, L} \rbrace,\lbrace  b_{23}, c_2, a_{12} \rbrace, R, \textcolor{blue}{I}; \lbrace \textcolor{blue}{L} \rbrace )$\\
\end{center}
where $\lbrace \textcolor{blue}{I, J, K, L} \rbrace$ are the states,\\
$\lbrace b_{23}, c_2, a_{12} \rbrace,$ are the input alphabets,\\
{R} is the set of rules,\\
$\textcolor{blue}{I}$ is a start state, and\\
$\lbrace \textcolor{blue}{L} \rbrace$ is a final state.\\

\noindent \textbf{with}

\begin{center}
	\textbf{R} =  $\lbrace \textcolor{blue}{I}b_{23} \rightarrow \textcolor{blue}{J}, \textcolor{blue}{J}c_{2} \rightarrow \textcolor{blue}{K}, \textcolor{blue}{K}a_{12} \rightarrow \textcolor{blue}{L}\rbrace$\\
\end{center}

\noindent \textbf{accepts} 

\begin{center}
	L(M) = $ \lbrace w \in \lbrace b_{23}, c_2, a_{12} \rbrace$*$: |b_{23}| = |c_{2}| = |a_{23}| \rbrace$\\
	i.e. $ b_{23} =  {al} ; c_2 = customers ; a_{12} =  vind $  \\
\end{center}

\begin{align*}
b_{23}c_2a_{12}a_{12}c_2\textcolor{blue}{\underline{I}}b_{23}  &\curvearrowright b_{23}c_2a_{12}\textcolor{blue}{\underline{J}}a_{12}c_2 &[\textcolor{blue}{I}b_{23} \rightarrow \textcolor{blue}{J}] \\
&\curvearrowright b_{23}c_2a_{12}\textcolor{blue}{\underline{K}}a_{12} &[\textcolor{blue}{J}c_{2} \rightarrow \textcolor{blue}{K}] \\
&\curvearrowright \textcolor{blue}{\underline{L}}b_{23}c_2a_{12} &[\textcolor{blue}{K}a_{12} \rightarrow \textcolor{blue}{L}] 
\end{align*}

\begin{figure}[H]
	\centering
	\includegraphics[width=0.95\linewidth]{1.png}
		\caption{An example of a sample JFA} \label{fig:M1}
\end{figure}

Figure \ref{fig:M1} shows an example of a sample JFA with three transitions. The diagram consists of four states, labelled as $\lbrace \textcolor{blue}{I, J, K, L} \rbrace$, where the start state is $\textcolor{blue}I$ and accepting state is $\textcolor{blue}L$. \textcolor{blue}{I}$b_{21}$ begins the movement with the \textit{\textcolor{bostonuniversityred}{al}} keyword, then \textcolor{blue}{J}$c_{2}$ shows the movement with the \textit{\textcolor{blue}{customer}} keyword, and the transition, \textcolor{blue}{K}$a_{12}$ ends with the \textit{\textcolor{ao(english)}{vind}} keyword. Since local languages are rich in vocabulary, the input alphabet may differ in another language, which might result in a change in the states and transitions.

\medskip
\begin{mdframed}[backgroundcolor=cornsilk!20]	
	\begin{itemize}
		\item Ngifuna \textcolor{ao(english)}{ukuthola} \textcolor{bostonuniversityred}{yonke} imininingwane ya ma \textcolor{blue}{customer}
	\end{itemize}
\end{mdframed} 

\noindent The above sentence is the Zulu translation of: ``I want to \textcolor{ao(english)}{find} \textcolor{bostonuniversityred}{all} the \textcolor{blue}{customer}'s details". The equivalent JFA shows:

\begin{center}
	\textbf{M} = $( \lbrace \textcolor{blue}{I, J, K, L} \rbrace,\lbrace a_3, b_{24}, c_2 \rbrace, R, \textcolor{blue}{I}; \lbrace \textcolor{blue}{L} \rbrace )$\\
\end{center}
where $\lbrace \textcolor{blue}{I, J, K, L} \rbrace$ are the states,\\
$\lbrace a_3, b_{24}, c_2 \rbrace,$ are the input alphabets,\\
{R} is the set of rules,\\
$\textcolor{blue}{I}$ is a start state, and\\
$\lbrace \textcolor{blue}{L} \rbrace$ is a final state.\\

\noindent \textbf{with}

\begin{center}
	\textbf{R} =  $\lbrace \textcolor{blue}{I}a_3 \rightarrow \textcolor{blue}{J}, \textcolor{blue}{J}b_{24} \rightarrow \textcolor{blue}{K}, \textcolor{blue}{K}c_2 \rightarrow \textcolor{blue}{L}\rbrace$\\
\end{center}

\noindent \textbf{accepts} 

\begin{center}
	L(M) = $ \lbrace w \in \lbrace a_3, b_{24}, c_2 \rbrace$*$: |a_3| = |b_{24}| = |c_2| \rbrace$\\
	i.e. $a_3 =  ukuthola ; b_{24} =  {yonke} ; c_2 = customers$ \\
\end{center}

\begin{align*}
b_{24}a_{3}c_2b_{24}c_2\textcolor{blue}{\underline{I}}a_3  &\curvearrowright b_{24}a_3c_2\textcolor{blue}{\underline{J}}b_{24}c_2 &[\textcolor{blue}{I}a_{3} \rightarrow \textcolor{blue}{J}] \\
&\curvearrowright b_{24}a_3c_2\textcolor{blue}{\underline{K}}c_2 &[\textcolor{blue}{J}b_{24} \rightarrow \textcolor{blue}{K}] \\
&\curvearrowright \textcolor{blue}{\underline{K}}b_{24}a_3c_2 &[\textcolor{blue}{K}c_2 \rightarrow \textcolor{blue}{L}] 
\end{align*}

\begin{figure}[H]
	\centering
	\includegraphics[width=0.95\linewidth]{2.png}
	\caption{An example of a sample JFA} \label{fig:m2}
\end{figure}

Figure \ref{fig:m2} shows an example of a JFA with three transitions. The four states as shown in the diagram are $\lbrace \textcolor{blue}{I, J, K, L} \rbrace$, with $\textcolor{blue}I$ representing the start state and $\textcolor{blue}L$ denoting the accepting state. \textcolor{blue}{I}$a_{3}$ begins the movement with the \textit{\textcolor{ao(english)}{ukuthola}} keyword, then \textcolor{blue}{J}$b_{24}$ shows the next input label with the \textit{\textcolor{bostonuniversityred}{yonke}} keyword, \textcolor{blue}{K}$c_{2}$ ends with the \textit{\textcolor{blue}{customer}} keyword. It is interesting to note that these examples only cater for a single relation. We may have situations of many states and transitions, taking into account the input statement provided by the learner.

\subsection{Normalising Queries}
At the normalisation phase, irrelevant details were removed and keywords that appears in our JFA design were used to semantically form a query. The normalisation process is passed to the SQL query generator, which is then used to create and retrieve records from the Chinook DB. As illustrated in Table \ref{table2}, each local language word is mapped to a query operation. This stage is useful for the SQL generation phase.

\begin{center}
	\captionof{table}{Mapping Local Language Keywords}
\label{table2}
\begin{tabular}{|l|l|}
	\hline
	\textbf{Keywords}                                                                                                        & \textbf{Query Operation} \\ \hline
	{\color[HTML]{009901} \begin{tabular}[c]{@{}l@{}}Ukuthola, Thola, Ngtholele, \\ Fumana, Ngitholela,\\ Vind, Kies\end{tabular}} & SELECT               \\ \hline
	{\color[HTML]{FE0000} \begin{tabular}[c]{@{}l@{}}Al, Alle, Alles, Konke, \\Yonke, Yothe, Vhothe\end{tabular}}              & ALL                  \\ \hline
	{\color[HTML]{009901} Skep, Usika, Dala}                                                                                   & CREATE               \\ \hline
	{\color[HTML]{009901} Faka, Plaas, Ulonga}                                                                                 & INSERT               \\ \hline
	{\color[HTML]{009901} \begin{tabular}[c]{@{}l@{}}Susa, Lees, Verywyder,\\ Utomola, Ubvisa Vhothe\end{tabular}}                                                             & DELETE               \\ \hline
\end{tabular}
\end{center}

\subsection{Generating SQL Queries}
This phase takes information from the normalisation phase to semantically generate a query. The generated query is used against the Chinook DB, which then produces an output. Since this task is limited to a single relation, the JFA algorithm was useful for this purpose. Similarly, the task was quite straightforward.

\begin{figure*}
	\centering
	\includegraphics[width=0.80\linewidth]{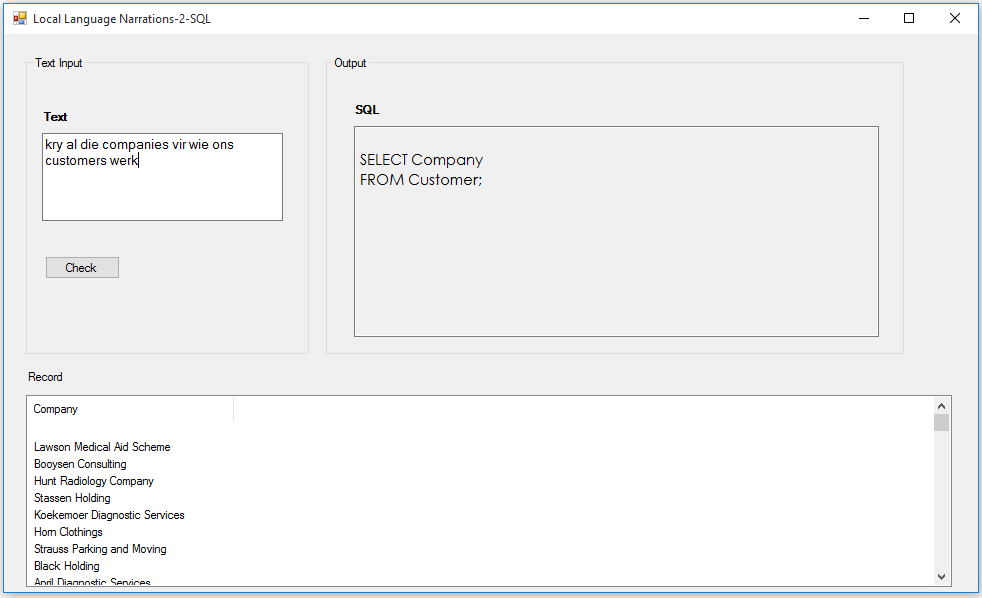}
	\caption{A sample output of \texttt{Local-Nar-SQL}}
	\label{fig:output}
\end{figure*}

\section{IMPLEMENTATION AND RESULTS}\label{results}
In Section \ref{methodology}, JFA technique was described which showed how a query is generated. Here, the implementation and results of the local language narration process are presented. The JFA approach was implemented using the Microsoft C$\#$
.NET framework\footnote{https://docs.microsoft.com/en-us/dotnet/framework/}, which was developed into a software tool, called \texttt{Local-Nar-SQL}. 

\texttt{Local-Nar-SQL} was created to allow a native learner to specify a query request in any of the eleven local languages spoken in SA. The software tool uses the learner's input to generate a query and retrieve records from the sample database.  Further, the result is made available to the learner. The sample database used for our use-case was the ChinookDB\footnote{https://github.com/lerocha/chinook-database}, which contains multiple records. A sample output of this request is provided in Figure \ref{fig:output}.

\section{EVALUATION}\label{evaluation}
\begin{figure*}[htb]%
	\centering
	\subfigure[Familiarity with SQL][Familiarity with SQL]{%
		\label{fig:ex3-a}%
		\includegraphics[height=1.4in]{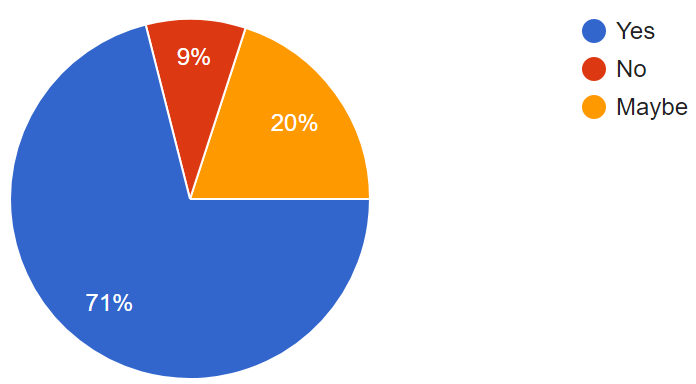}}%
	\hspace{8pt}%
	\subfigure[Proficiency in a local language][Proficiency in a local language]{%
		\label{fig:ex3-b}%
		\includegraphics[height=1.5in]{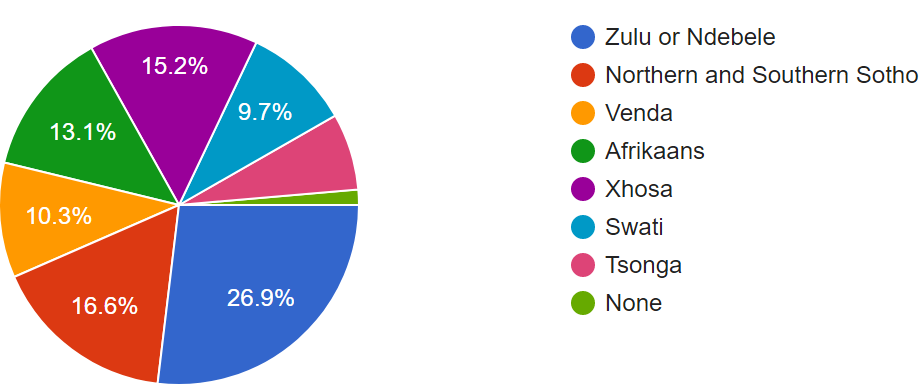}} \\
	\subfigure[A correct translation of a local language narrations][A correct translation of a local language narrations]{%
		\label{fig:ex3-c}%
		\includegraphics[height=1.5in]{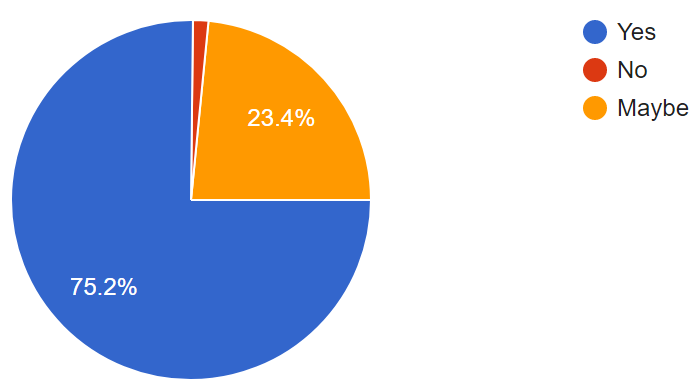}}%
	\hspace{8pt}%
	\subfigure[Assist native learners work with SQL][Assist native learners work with SQL]{%
		\label{fig:ex3-d}%
		\includegraphics[height=1.5in]{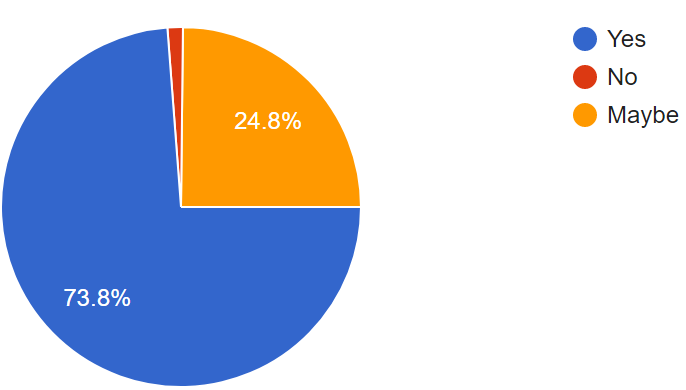}}%
	\caption[The survey results]{The survey results}
	\label{fig:ex3}
\end{figure*}

A total of 145 participants participated in a survey that was conducted online. The majority of the participants are native language learners who are familiar with programming and have taken numerous computer science courses. The questions for the survey is available via \url{https://forms.gle/uSXn8TZpBEWmnGk66}. The results are presented in Figure \ref{fig:ex3}.

Of the responses received, 71\% indicated familiarity with SQL, whilst 9\% of respondents claimed no familiarity with SQL and 20\% were unsure of their answers (Figure \ref{fig:ex3-a}). The participants were asked about their native language proficiencies. Approximately 29\% of learners were proficient in Zulu, 16.6\% were proficient in Northern and Southern Sotho, 15.2\% claimed to be proficient in Xhosa and 13.1\% admitted to being proficient in Afrikaans. The remaining respondents were proficient in Venda (10.3\%), Swati (9.7\%) or Tsonga (6.9\%), and 1.4\% of the participants did not list any of these native languages (Figure \ref{fig:ex3-b}). Furthermore, the participants were asked if the \texttt{Local-Nar-SQL} tool provided a correct translation of local language narrative to SQL: 75.2\% affirmed that the tool was accurate, 23.4\% were unsure and 1.4\% indicated no (Figure \ref{fig:ex3-c}). When the participants were asked if they think that the tool will assist native learners work with SQL, about 73.8\% affirmed that the tool would assist learners, 24.8\% were unsure and 1.4\% indicated no (Figure \ref{fig:ex3-d}).  

The survey showed that the majority of the learners found the \texttt{Local-Nar-SQL} tool to be useful and helpful towards their interest of learning SQL. If our proposed tool is used as a learning aid, native learners will find it easier to work with SQL, which will eliminate many of the barriers faced with English proficiencies in programming pedagogies. 

\section{CONCLUSION AND FUTURE WORK}\label{conclusion}
This paper presented an extended version of a JFA for the automatic synthesis of SQL queries from local language narrations. This automata-based algorithm was further implemented into a software tool called \texttt{Local-Nar-SQL}, designed for native learners to specify query request in any of the eleven official South African languages of their choice.  \texttt{Local-Nar-SQL} takes these requests, generates a query and produces an output that is presented to the learner. If implemented on a large scale, this tool will serve as:

\begin{enumerate}
 \item a comprehension aid: With \texttt{Local-Nar-SQL}'s textual and translation approaches, it can improve the cognitive workload of a native learner towards understanding the SQL concept.
 
 \item a learning aid: The \textit{free-form} approach offered by the tool will enable native learners improve and enhance their SQL query skills.   
 
 \item a solution to English language barriers: Native learners are not required to attain a high English proficiency to use \texttt{Local-Nar-SQL}. Since local languages are widely spoken in South Africa, it is seen that this tool will stimulate a native learner's interests in becoming proficient in SQL. 
\end{enumerate}
To this end, our evaluation results showed that the majority of learners agreed that this tool will be helpful to native learners yearning to be proficient in SQL. Our evaluation results are consistent with the works of \cite{siciliani2019italian} where majority of the native participants agreed that the developed tool helped serve their information needs . In the future, we aim to improve this study to accommodate nested SQL query and other query operation tasks. Also, it would be interesting to conduct a performance evaluation on our proposed tool. Such information would provide us with the accuracy of \texttt{Local-Nar-SQL}.

\paragraph{Conflict of Interest}

The authors declare no conflict of interest.

\paragraph{Acknowledgment}
This work is based on research supported by the National Research Foundation (NRF) of South Africa (Grant Number: 119041). Any opinion, findings and conclusions or recommendations expressed in this material are those of the authors and therefore the NRF does not accept liability in regard thereto.

\bibliographystyle{refastesj}
\bibliography{references}  
\end{multicols}
\end{document}